# Interpretation and Analysis of the Steady-State Neural Response to Complex Sequential Structures: a Methodological Note

Nai Ding

College of Biomedical Engineering and Instrument Science,

Zhejiang University, Hangzhou, China

**Abstract**

Frequency tagging is a powerful approach to investigate the neural processing of sensory features, and is recently adapted to study the neural correlates of superordinate structures, i.e., chunks, in complex sequences such as speech and music. The nesting of sequence structures, the necessity to control the periodicity in sensory features, and the low-frequency nature of sequence structures pose new challenges for data analysis and interpretation. Here, I discuss how to interpret the frequency of a sequential structure, and factors that need to be considered when analyzing the periodicity in a signal. Finally, a safe procedure is recommended for the analysis of frequency-tagged responses.

# 1. Introduction

Frequency tagging is a power technique to extract the neural response tracking a stimulus feature. In general, in the frequency tagging paradigm, a target stimulus feature is periodically modulated at a frequency $f$. Consequently, the neural response that dynamically tracks the stimulus feature also fluctuates at frequency $f$. The $f$-Hz frequency tagged response is often extracted using the Discrete Fourier Transform (DFT) or wavelet transform. Frequency-tagging is a powerful paradigm for electroencephalography (EEG) and magnetoencephalography (MEG) studies since it can extract any neural response that follows the $f$-Hz change in the stimulus, regardless of the latency or waveform of the response. The paradigm has been widely applied to study visual (Norcia et al., 2015; Regan, 1977) and auditory (Galambos et al., 1981; Picton et al., 2003) processing: The frequency-tagged response to periodic changes in visual features, e.g., luminance, is referred to as the Steady State Visual Evoked Potentials (SSVEP), while the frequency-tagged response to periodic changes in auditory features, e.g., intensity, is referred to as the auditory Steady State Response (aSSR). These responses are widely applied to study the basic properties of sensory encoding (Herrmann, 2001; Ross et al., 2000; Wang et al., 2012; Wong et al., 2007) and cognitive control (Andersen et al., 2008; Elhilali et al., 2009; Gao et al., 2021).

More recently, the frequency-tagging paradigm has been applied to study the neural processing of superordinate structures in complex sequences, e.g., speech and music: The hypothesis in these studies is that a mentally constructed superordinate sequence structure, i.e., a sentence, is neurally represented by a response whose duration matches the duration of the structure in the stimulus (Buiatti et al., 2009; Ding et al., 2016; Nozaradan et al., 2011). On the one hand, frequency tagging provides a powerful paradigm to investigate the neural processing of a chunk in contrast to a brief stimulus event and has stimulates a large number of studies (Batterink & Paller, 2019; Benitez-Burraco & Murphy, 2019; Choi et al., 2020; Glushko et al., 2022; Henin et al., 2021; Kaufeld et al., 2020; Kazanina & Tavano, 2022; Keitel et al., 2018; Lo et al., 2022; Lu et al., 2021; Makov et al., 2017; Meng et al., 2021; Meyer, 2018). On the other hand, the complexity of the sequence processing problem has also caused more challenges to the analysis and interpretation of the frequency-tagged responses. First, in traditional frequency tagging studies, each stimulus feature of interest is tagged at a distinct frequency, while the structures in a complex sequence are often nested so that different levels of structures cannot be tagged at unrelated arbitrary frequencies. For example, in a sentence "the cute boy smiled", the first three words construct a noun phrase based on syntax. Nevertheless, the 3-word noun phrase and the 4-word sentence are nested so that they cannot be frequency tagged at unrelated frequencies. The nesting between structures lead to a dissociation between structure duration and structure repetition period, which is discussed in Section 2.1.

Second, traditional frequency tagging studies explicitly create periodic changes in a stimulus feature while the studies on sequence structures sometimes want to avoid such periodic changes in basic stimulus features to isolate the neural response generated by internal mental processes. What is a neural response generated by internal mental processes? For example, a metrical structure may be imagined when listening to an isochronous beat sequence, and the neural response at the imagined meter rate can reflect internally driven processes (Nozaradan et al., 2011). Similarly, when a sequence of words is grouped into sentences based on syntactic rules, the neural response at the sentence rate can reflect higher-level sentence processing (Ding et al., 2016). In these situations, however, if a basic sensory feature has the same periodicity as the imagined meter or syntactically constructed sentence, it is ambiguous whether the neural response tracks the sensory feature or the sequence structures. Therefore, it is often necessary to check the periodicity in stimulus features. Cautions, however, are needed since some types of periodicities are not captured by the Fourier transform, which is discussed in Section 2.2.

Third, the analysis of responses to frequency-tagged sequence structures is sometimes prone to artifacts that seldomly affect the analysis of traditional frequency tagged responses. For sequence structures often correspond to a very low frequency, e.g., < 3 Hz, and such a low-frequency may be contaminated by overlapping in the analysis epochs (Benjamin et al., 2021). Section 3 illustrates why such artifacts may be generated and discusses potential guidelines for appropriate analysis of the frequency tagged responses, including the selection of analysis duration and whether a smoothing window should be used. This article discusses common technical issues, instead of the analysis of a specific experiment. However, to facilitate interpretation, a hypothetical experiment is provided in Fig. 1A, but the conclusions are not limited to this example. On the other hand, the target audience is experimentalists instead of engineers. Therefore, article attempts to explain ideas using illustrations and skips mathematical derivations. The mathematical basis of the DFT can be found in classic textbooks such as Oppenheim et al. (2001).

## 2. What is not reflected by the Fourier transform
### 2.1 Frequency may not reflect the time constant or signal duration

For frequency-domain analysis, a central concept is frequency, which corresponds to the period of a signal. The period of a signal, however, does not necessarily coincide with other time constants of a signal. For example, an exponential signal $e^{-t/\tau}$ has a time constant $\tau$, but the signal is aperiodic and $\tau$ is not a period of the signal. Even for a periodic signal, its period may dissociate from the time constant or duration of the waveform within a period, and some examples are shown in Fig. 1B. In these examples, the signals have a period of 1 s, and the waveform within a period is shown in the left panel. The temporal parameters of the signal, including the time constant of an exponential function, duration of a sawtooth signal, and frequency of a single-cycle sinusoid, affect the shape of the Fourier spectrum but generally do not lead to any spectral peak corresponding to these parameters. Instead, since the signal repeats at a rate of 1 Hz, the spectrum shows peaks at 1 Hz and its harmonically related frequencies, i.e., 2 Hz, 3 Hz, etc. When the period of the signal changes, however, the spectral peaks shift accordingly, even if the waveform within a cycle remains unchanged (Fig. 1C). See Zhou et al. (Zhou et al., 2016) for more illustrations about how the spectrum is influenced by the signal repetition rate and the waveform within a period.

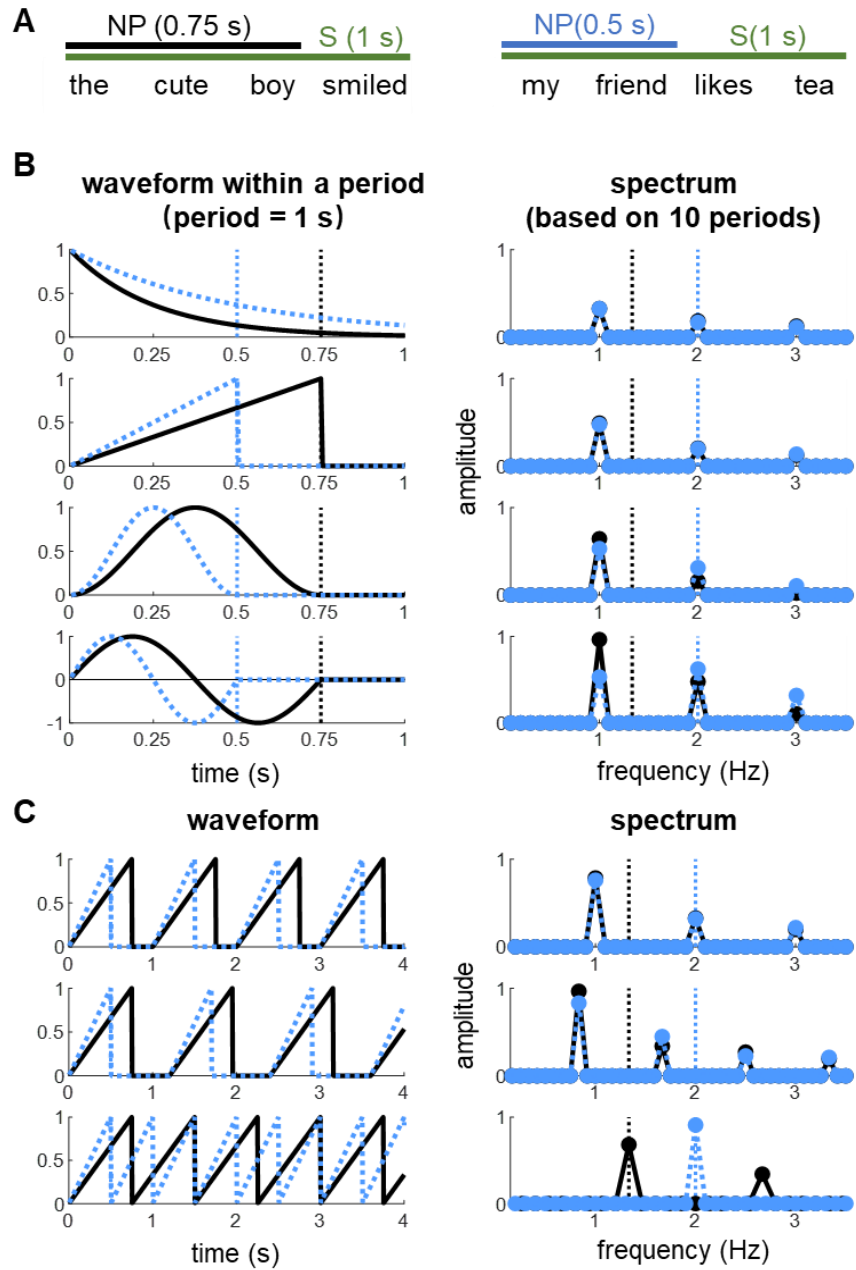

Figure 1. Peaks in the spectrum reflects the periodicity of a signal. A) A hypothetical experiment condition, in which a noun phrase (NP) is embedded in a sentence (S). The duration of the NP is either 0.75 s or 0.5 s, and a neural response is hypothesized to be modulated the duration of the NP. B) Signals that repeat every 1 s and the corresponding spectra. The left panel shows the waveform within a period, and the black and blue curves have different time constants, i.e., 0.75 s and 0.5 s respectively. The right panel shows the spectrum that is the magnitude of the DFT transform of 10 periods of the corresponding signal. The time constant and the corresponding frequency are shown by the vertical dotted lines. The spectrum has peaks at 1 Hz, 1 over the signal period, and harmonically related frequencies, regardless of the time constant of the signal within a period. C) Signals that are constructed by the same sawtooth waveform but have different repetition rates. The spectral peaks always reflect the repetition rate.

## 1.2 Frequency may not reflect the rate of change

Suppose a signal changes every $T$ s. Intuitively, its Fourier spectrum should peak at $1/T$ Hz. This intuition, however, is not always true and an example is given in Fig. 2, in which the spectrum shows troughs at $1/T$ Hz and harmonically related frequencies. When the signal is employed to modulate the gain of a 4-Hz sinusoid, the modulated sinusoid does not show any power at $1/T$ Hz either. The purpose of these examples is to show that the Fourier transform may be blind to some rhythms. Why does the signal lack power at $1/T$ Hz? In the Fourier transform, the power at $f$ is determined by the dot product between the signal and sinusoids at frequency $f$ (including both sine and cosine). The signals in Fig. 2 contain no fluctuations within each $T$ s and therefore the signal has no correlation with sinusoids at $1/T$ Hz. Figure 3 illustrates the dot product between signals.

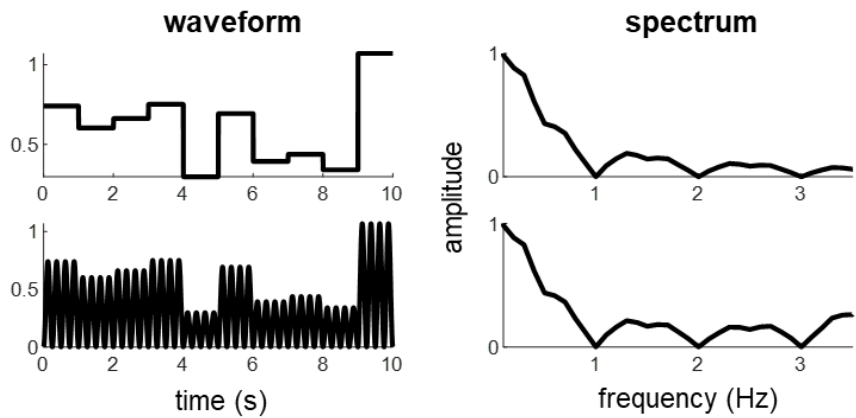

Figure 2. The change rate of a signal can correspond to troughs in the spectrum. The upper panel shows a signal that changes once every 1 s, and the lower panel is a 4-Hz sinusoid that is amplitude modulated by the signal on the upper panel. In the spectrum, troughs are observed at 1 Hz and harmonically related frequencies.

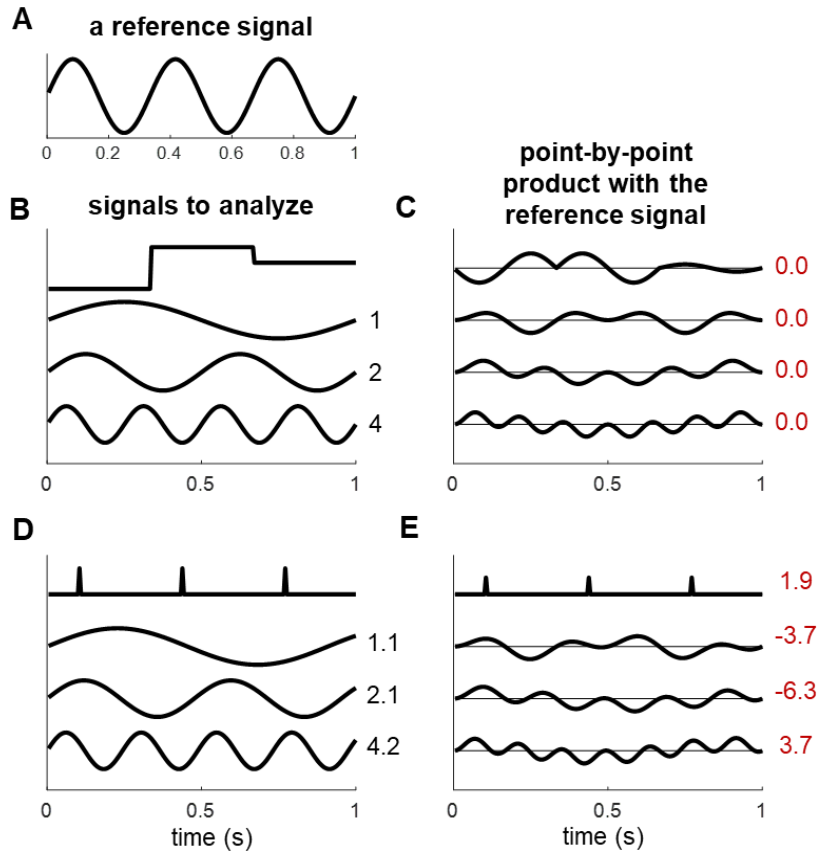

Figure 3. Illustration of the dot product between signals, which is the basis of the DFT. A) A 3-Hz sinusoid, which is employed to calculated the DFT coefficient at 3 Hz. BC) Signals to analyze and their point-by-point product with the reference signal. The top signal is similar to the signal in Fig. 2, while the other 3 signals are sinusoids with the frequency shown by the number in panel B. The sum of the product signal, i.e., the dot product between the two signals, is shown by the number in red in panel C. DE) Examples of signals that have nonzero dot product with the reference signal.

## 2. Effects of the neural response analysis method

### 2.1 Overlapping epochs can introduce artifacts

A rhythm can be created based on an arbitrary signal by adding delayed versions of the signal to itself. An illustration is shown in Fig. 4A, in which the signal to analyze only consists of a pulse at 4.8 s and is 0 otherwise. When the signal is chunked into 5-s epochs with 4-s overlap, however, the averaged epoch clearly becomes periodic and the period is the same as the distance between adjacent epochs, e.g., 1 s. Another example is shown in Fig. 4B, in which a white noise is chunked into 5 s epochs in the same way. The spectrum averaged over 100 epochs clearly shows a peak at 1 Hz. In fact, in hearing research, this method has been employed to generate pitch perception based on, e.g., white noise (Yost, 1996).

If inappropriate data epoching can introduce artifacts, why not directly applying the Fourier transform to the unepoched data? A direct Fourier transform to the unepoched data can indeed yield a high-frequency-resolution spectrum of the response. Nevertheless, in real EEG/MEG recordings, strong artifacts caused by, e.g., head movements or hardware glitches, can barely be avoided during a long recording, and excluding recordings with large artifacts from further analyses is a common practice in EEG/MEG analysis. It is nonoptimal, however, to throw away a long recording based on a few sparsely located artifacts. Therefore, segmenting a long recording into shorter epochs and only removing epochs with obvious artifacts is a common strategy

## 2.2 Analysis window determines the width of spectral peaks

Suppose a frequency-tagged neural response has a period of $T$ s, and $D$ seconds of recording is transformed into the frequency domain using the DFT. The DFT spectrum consists of coefficients corresponding to discrete frequencies, i.e., $1/D$ Hz, $2/D$ Hz, $3/D$ Hz, etc. If $D$ is a multiple of $T$, the frequency-tagged response is resolved in the spectrum. In other words, if $D = kT$, where $k$ is an integer, the $k^{\text{th}}$ DFT coefficient corresponds to $1/T$ Hz, i.e., the target frequency. In this case, the response spectrum only has power at $1/T$ Hz and harmonically related frequencies. An example is shown in Fig. 5A (upper panel), where $T$ is 0.5 s, $D$ is 5 s, and the neural response is exactly a sinusoid. The response spectrum has a sharp peak at 4 Hz and the power in adjacent frequency bins is 0. The DFT coefficient not at 4 Hz is zero since the dot product between any two $D$-s long sinusoids at frequencies resolved by the DFT is zero (Fig. 3BC). When $D$ is not a multiple of $T$, however, the DFT spectrum does not have a frequency bin corresponding to $1/T$ Hz and the power of the signal spreads to many frequency bins near $1/T$ Hz, a phenomenon known as frequency leakage. An example is shown Fig. 5B (upper panel), where $T$ is still 0.5 s but $D$ is 5.1 s.
.

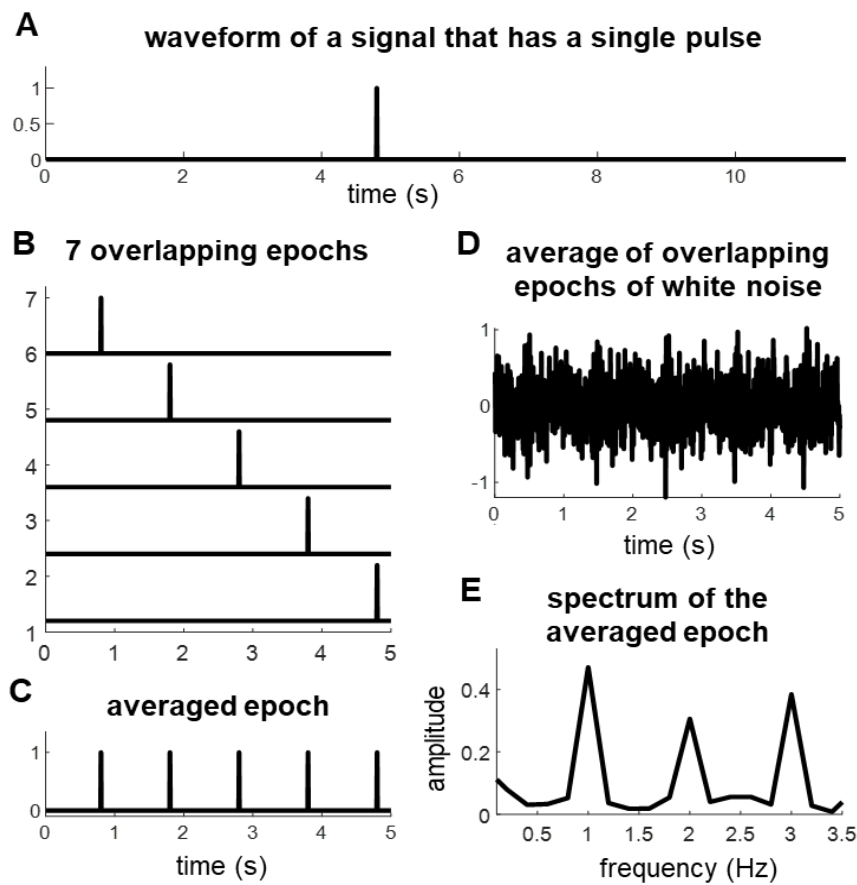

Figure 4. Overlapping epochs can lead to spurious peaks in the spectrum. A) A nonperiodic signal that is composed of a single pulse. B) The signal in A is segmented into 5-s epochs that have 4-s overlap with each other. C) The average of the epochs in B. D) The same epoching process is applied to white noise and the resulting waveform is shown. E) The spectrum of the signal in D. To obtain a robust result, twenty independent white noise is generated and processed the same way and the spectra are averaged.

A common strategy to alleviate frequency leakage is to multiply a smoothing window to the signal before the Fourier transform. The spectra of the windowed signals are shown in Fig. 5 (lower panel). With the smoothing window, the signal duration no longer strongly affects the shape of the spectrum, but the spectrum always has nonzero power in frequency bins near the target frequency, i.e., 2 Hz. The main difference between the methods in Fig. 5 is whether all the power of a sinusoid concentrates in a single frequency bin or spreads to several bins. It is not further illustrated but the conclusions apply to other variations in the analysis method, such as padding zeros to the signal or using the wavelet transform instead Fourier transform.

Shall we care about whether the signal power concentrates in a single frequency bin or not? The answer is yes in conditions. For example, a convenient approach to test the statistical significance of a frequency tagged response is to compare the power at the target frequency with the power in adjacent frequency bins (Benjamin et al., 2021; Ding et al., 2016; Nozaradan et al., 2011). The statistical power of this approach is clearly compromised when the power in adjacent frequency bins is elevated from baseline. Even when the statistical significance of the frequency-tagged response is tested using other methods, e.g., in comparison with a control condition that does not have the frequency-tagged response (Andersen et al., 2008), the statistical power of the test can benefit from concentrating all power of the frequency-tagged response into a single frequency.

More generally, when the periodicity of a signal is unknown and needs to be determined using the Fourier analysis, a smoothing window often helps. Nevertheless, in the frequency-tagging paradigm, the target frequency is known and therefore a smoothing window is not necessary. In other words, in the frequency-tagging approach, the purpose of data analysis is not to estimate the periodicity of a response but to detect the presence of a response with a known frequency. Based on the signal detection theory (Poor, 1998), optimal detection of a sinusoid generally involves calculating the dot product between the recorded signal and the target signal, which can be viewed a sinusoid in the frequency tagging paradigm, and such dot product can be conveniently calculated using the DFT.

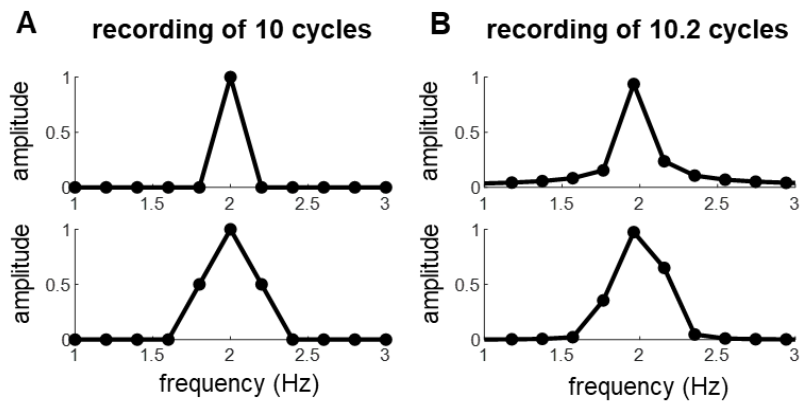

Figure 5. Frequency leakage and windowing. The signal to analyze is a 2-Hz sinusoid and the duration of the signal is 5 s in panel A and 5.1 s in panel B. The upper panel is the DFT of the signal and the lower panel is the DFT of the signal smoothed by a Hanning window.

## 3. Summary

First, in the frequency tagging paradigm, the target frequency is the frequency at which a stimulus feature or sequence structure repeats, which in general does not relate to how long the feature or structure lasts or how fast it varies within each period. Second, the Fourier transform does not provide a one-size-fits-all solution to extract all periodicities in a signal. On the stimulus side, cautions are needed, e.g., when making sure that a stimulus does not contain any conceivable periodicity at a target frequency. On the response side, more advanced feature extraction methods may be necessary to identify a frequency-tagged response. For example, for the signals in Fig. 2, taking the absolute value of the first-order derivative of the signal can reflect the 1-Hz periodicity in the signal.

Finally, I recommend the following as a relatively safe procedure to analyze frequency-tagged responses.

(1) The response being analyzed should contain exactly an integer number of periods of the frequency-tagged response. More specifically, if the response sampling rate is $F$ and the response is frequency tagged at $f$, the number of samples per cycle of the response is $F/f$, which does not need to be an integer. Nevertheless, if $k$ cycles are included in the analysis window, the total length of the analysis window, i.e., $kF/f$, should be an integer and $k$ is also an integer.

(2) When the stimulus lasts for a very long duration (e.g., several minutes), and the response recorded throughout the presentation of the stimulus can be directly transform into the frequency domain. Alternatively, it can be segmented into shorter epochs, e.g., to remove epochs with large artifacts, and averaged. The epochs, however, should not overlap.

(3) No smoothing window is necessary when performing the Fourier analysis, when the target frequency is known and the analysis window contains an integer number of cycles of the target response.

**Acknowledgement**

I thank Wenhui Sun for helping formatting the bibliography. This work was supported by the National Natural Science Foundation of China (32222035) and Key R & D Program of Zhejiang (2022C03011).